\documentclass[aps,prl,twocolumn,showpacs,superscriptaddress,groupedaddress]{revtex4}  
\usepackage{graphicx}  
\usepackage{dcolumn}   
\usepackage{bm}        
\usepackage{amssymb} 
\usepackage{amsmath}   
\usepackage{braket}
\begin{document}



\title{Operating Quantum States in Single Magnetic Molecules: Implementation of Grover's Quantum Algorithm}

\author{C. Godfrin}
\affiliation{CNRS Institut N\'eel, Grenoble, F-38000, France \\}
\affiliation{Univiversit\'e Grenoble Alpes, Institut NEEL, Grenoble, F-38000, France \\}

\author{A. Ferhat}
\affiliation{CNRS Institut N\'eel, Grenoble, F-38000, France \\}
\affiliation{Univiversit\'e Grenoble Alpes, Institut NEEL, Grenoble, F-38000, France \\}

\author{R. Ballou}
\affiliation{CNRS Institut N\'eel, Grenoble, F-38000, France \\}

\author{S. Klyatskaya}
\affiliation{Institute of Nanotechnology, Karlsruhe Institute of Technology, 76344 Eggenstein-Leopoldshafen, Germany\\}

\author{M. Ruben}
\affiliation{Institute of Nanotechnology, Karlsruhe Institute of Technology, 76344 Eggenstein-Leopoldshafen, Germany\\}

\author{W. Wernsdorfer}
\affiliation{CNRS Institut N\'eel, Grenoble, F-38000, France \\}
\affiliation{Physikalisches Institut, Karlsruhe Institute of Technology, D-76131 Karlsruhe, Germany \\}

\author{F. Balestro}
\affiliation{CNRS Institut N\'eel, Grenoble, F-38000, France \\}
\affiliation{Univiversit\'e Grenoble Alpes, Institut NEEL, Grenoble, F-38000, France \\}
\affiliation{Inst. Univ. de France, 103 Blvd Saint-Michel, 75005 Paris, France}

\date{\today}

\begin{abstract}
	Quantum algorithms use the principles of quantum mechanics, as for example quantum superposition, in order to solve particular problems outperforming standard computation. They are developed for cryptography, searching, optimisation, simulation and solving large systems of linear equations. Here, we implement Grover'€™s quantum algorithm, proposed to find an element in an unsorted list, using a single nuclear 3/2-spin carried by a Single Molecular Magnet (SMM) transistor. The coherent manipulation of this multi-level qudit is achieved by means of electric fields only. Grover's search algorithm was implemented by constructing a quantum database via a multi-level Hadamard gate. The Grover sequence then allows us to select each state. The presented method is of universal character and can be implemented in any multi-level quantum system with non-equal spaced energy levels, opening the way to novel quantum search algorithms.
\end{abstract}

\maketitle

\section{I. Introduction}
A quantum algorithm is a finite succession of unitary transformations performed on an initially prepared quantum state that aims to generate a final quantum state encoding the answer to a problem. The concept has attracted considerable attention after the discovery in 1994 by P.W. Shor describing that such an algorithm can factorize an integer with an exponentially smaller number of operations than any known classical algorithm \cite{Shor1994}. A great deal of efforts is devoted since then to find new powerful quantum algorithms and to implement them in actual devices. Toward this goal, a variety of possible prototypes of a quantum bit (qubit) have been proposed \cite{Monroe1996, Brune1996, Nakamura1999, Kim2015, Jelezko2004} including nuclear spin systems \cite{Kane1998, Dutt2007, Neumann2010, Pla2013, Vincent2012}, and experimental proofs-of-concept of quantum algorithms have been worked out\cite{Lu2007, Lanyon2007, Politi2009, Martin2012, Cai2013, Barz2014, King2015, Jones1998, Chuang1998, Vandersypen2000}. A plethora of quantum algorithms has been formulated \cite{Jordan2017}, among which one distinguishes those specifically simulating quantum systems, initially suggested by R. Feynman \cite{Feynman1982}, and those relying on quantum Fourier transforms, such as the Shor's algorithm for integer factorization \cite{Shor1994}. The third main category of quantum algorithms incorporates the amplitude amplification discovered in 1997 by L. K. Grover to search an element in an unsorted list \cite{Grover1996, Grover1997}. Few years later, a theoretical study proposed the implementation of this algorithm using a molecular magnet \cite{Leuenberger2001}. The Grover algorithm first creates an initial equal superposition of states by means of a Hadamard gate. Then it iteratively applies a quantum oracle to negate the amplitude of the searched state followed by diffusion transform that inverts each amplitude about the average. Grover's algorithm was proven to be quadratically faster than any classical search algorithm: After a number of iterations close to the square root of the length of the database, the final state collapses onto the searched state with a high probability. In previous experiments, the proof-of-concept of the Grover's algorithm was demonstrated with nuclear magnetic resonance experiments on two \cite{Jones1998, Chuang1998} or three \cite{Vandersypen2000} entangled qubits, involving respectively 4 and 8 states, respectively. A fundamentally different approach to the Grover algorithm was proposed in 1998 that does not make use of entangled qubits nor of a quantum oracle \cite{Farhi1998}. Instead, it was proposed to implement the algorithm into multi-level systems and to proceed by controlled time evolution of the wave-functions of the different involved levels by driving Hamiltonians. Herein, we present the implementation of this approach using a single nuclear spin I=3/2 \cite{Leuenberger2003}. After describing this 4-level qudit, we present the coherent manipulation of each transition. Finally the experimental implementation of the multi-level Grover algorithm is described consisting of 2 subsequent quantum gate operations: (i) A Hadamard gate creates the quantum directory by a coherent superposition of all the states and (ii) the Hamiltonian of a unitary evolution will then make the system evolving to the desired state. \\

\begin{figure}
	\begin{center}
		\includegraphics[width=0.45\textwidth]{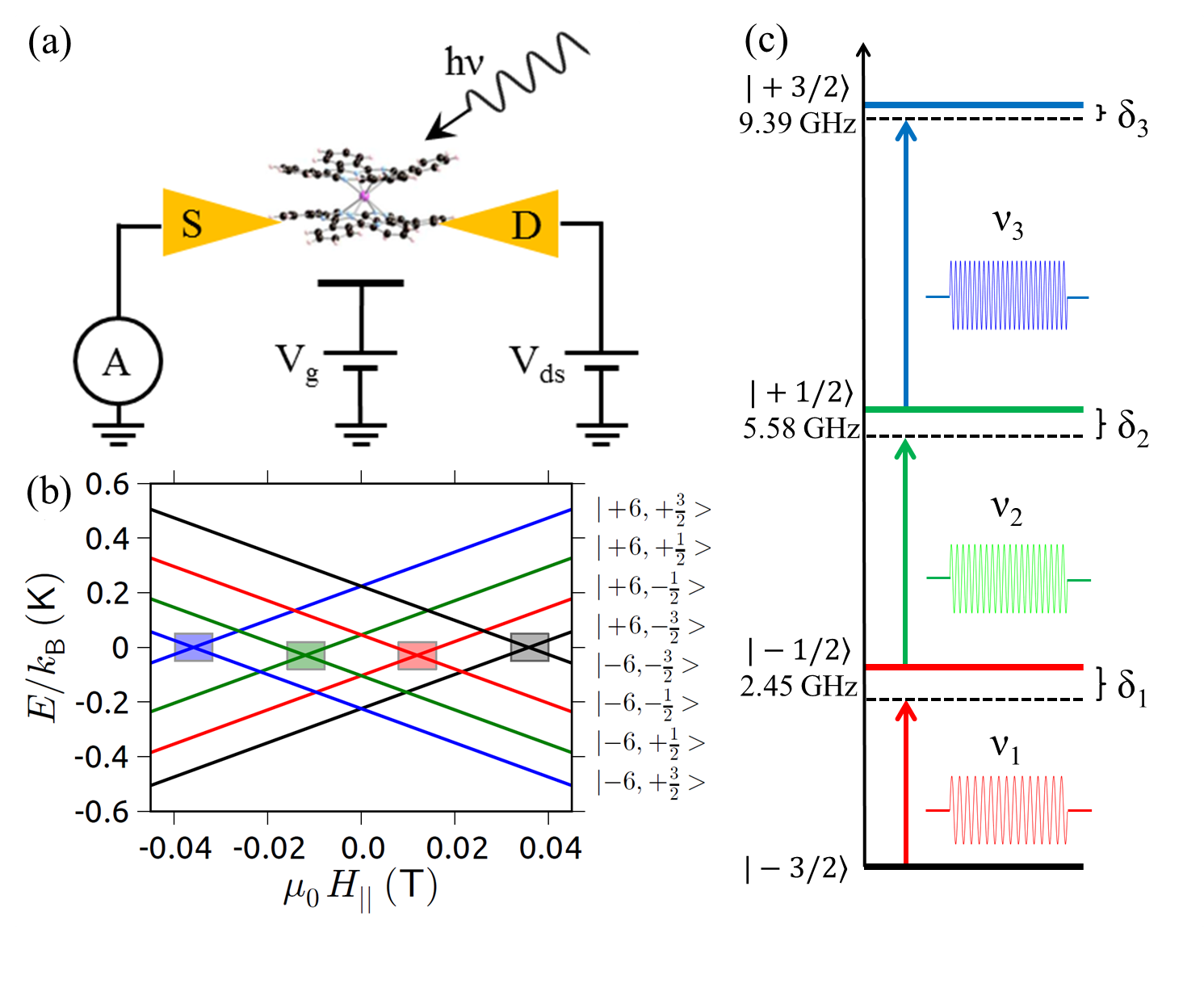}
		\caption{\label{fig1} (a) The TbPc$_2$ molecular magnet is embedded in a transistor, microwave pulses allow a coherent manipulation of the nuclear spin carried by the Tb$^{3+}$ ion. (b) Zeeman diagram of the molecule. The four colors square indicate the anti-crossing positions. By sweeping back and forth the magnetic field a QTM is possible at these four positions. The electronic spin conductance dependence of the transistor enables to read-out the nuclear spin states. (c)  Energy diagram of the four nuclear spin states. The quadrupole component in the hyperfine coupling enables an independent manipulation of each transition. We define the frequency of the microwave pulse $\nu_n$ that drives the $n-1  \leftrightarrow n$ transition and the detuning $\delta_n$ between the transition and microwave pulse frequency.}
	\end{center}
\end{figure}

\section{II. Reading-out nuclear spin states}
All experimental results presented in this work were obtained via electric transport measurements through a three terminal single-molecule magnet transistor. The device presented in Figure 1(a) consists of a bis(phthalocyanine)terbium (III) SMM (TbPc$_2$), contacted to two gold electrodes by using the electromigration technique \cite{Park2000}. The hearth of the molecule is a Tb$^{3+}$ ion that is eightfold N-coordinated by two phthalocyanine (Pc)-ligands. It exhibits an electronic configuration [Xe]4f$^8$ with a total spin S=3 and a total orbital momentum L=3. A strong spin-orbit coupling yields an electronic spin with a total angular magnetic moment J=6. In addition, the ligand field generated by the two Pc ligands leads to an energy gap of the order of 600K from the ground state doublet m$_J$=$\pm$6 compared to m$_J$=$\pm$5. Because the measurements are performed at very low temperature (40mK), the electronic spin can be considered as a $\pm$6 Ising spin with a uniaxial anisotropy axis perpendicular to the Pc-plane (Figure 1(a)). In addition to the electronic spin, the monoisotopically $^{159}$Tb$^{3+}$ ion carries a nuclear spin I=3/2. The hyperfine interaction of A$\approx$24.9mK \cite{Ishikawa2005} between the electronic and the nuclear spin results in a fourfold level splitting of each electronic spin state as presented in the Zeeman diagram (Figure 1(b)). The quadrupole term P$\approx$14.4mK of the hyperfine coupling yields a unequal energy level spacing between the four nuclear spin states, resulting in three different resonance frequencies $\nu_1\approx$2.45GHz, $\nu_2\approx$3.13GHz and $\nu_3\approx$3.81GHz (Figure 1(c)). Off-diagonal terms in the ligand-field Hamiltonian give rise to a finite tunnel probability from one electronic spin state into the other conserving the nuclear spin state. The coloured rectangles in Figure 1(b) indicate the position of the avoided level crossings where Quantum Tunnelling of the Magnetization (QTM) can occur. We previously reported that the magnetic moment both of the single electronic and nuclear spin can be read-out via transport measurements \cite{Vincent2012, Godfrin2017} and that coherent manipulation of a single nuclear spin can be performed using electric fields only \cite{Thiele2014}. In solid state devices, the read-out and coherent manipulation of a nuclear spin was also achieved for Nitrogen-Vacancy centers in diamond \cite{Neumann2010} and ionized $^{31}$P donor in Silicon \cite{Pla2013}. \\

\begin{figure}
	\begin{center}
		\includegraphics[width=0.45\textwidth]{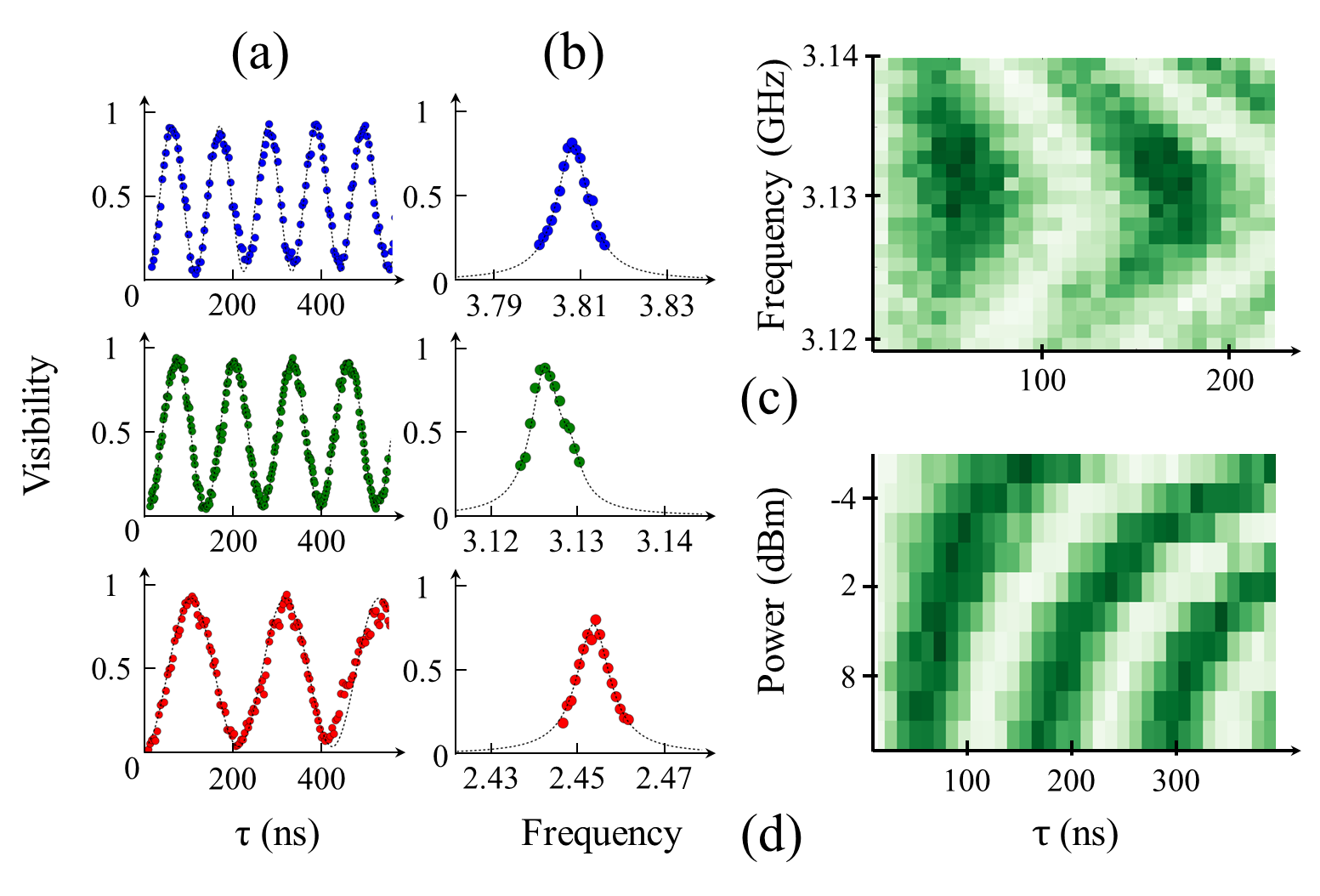}
		\caption{\label{fig2} (a) Rabi oscillations: the frequency of the oscillations can be tuned from 1.5MHz to 8MHz. Each colour represent a qubit transition:  1, 2 and 3 respectively in red, green and blue.(b) By recording the maximum of visibility of the Rabi oscillation as function of the detuning we measure the resonance shape of the three transitions. (c) Visibility of the second state as the function of the pulse length and frequency. The detuning increases the frequency of the oscillation and decreases the maximum of visibility. (d)  Visibility of the second state as the function of the pulse length and power. A linear dependence of the Rabi frequency as a function of the square root of the pulse power is measured. 
		}
	\end{center}
\end{figure}

The SMM transistor is cooled down using an inverse dilution refrigerator (electron temperature T$_{el}$=50mK) and subjected to static and low sweeping rate 3D-magnetic fields. Microwave pulses of frequency $\nu_{RF}$, amplitude $E_{RF}$ and duration $\tau_{RF}$ are applied via an antenna in the vicinity of the device using a monochromatic pulse synthetized by a Rhode \& Schwarz SMA100A generator with an AWG external pulse modulation. Repeated spin state initialization, coherent manipulation, and read-out use the following cycle: the external magnetic field is swept between $\pm$60 mT at 100 mT/s (Figure 3(a)) until a QTM transition is measured. Depending on the field value at which QTM occurs, the respective nuclear spin state can be accessed. Then, at constant external magnetic field, a microwave (MW) pulse is applied. Finally, the resulting state is detected by sweeping back the external magnetic field. The entire sequence is rejected when no QTM transition is detected. After repeating this procedure 1000 times for each pulse sequence, we yield the transition probability between the nuclear states i and j: 
\begin{equation}
P_{i,j}=\frac{N_{i,j}}{\sum_{n} N{i,n}}
\end{equation}
where N$_{i,j}$ is the number of events of initial state i and final states j. 
First, we study each nuclear spin transition separately. The transition Hamiltonian in the rotating frame is given by:

\begin{equation}
H_{qubit}=\pi \hbar (\delta \sigma_z + \Omega \sigma_x)=\pi \hbar \begin{pmatrix}
\delta & \Omega \\
\Omega & -\delta
\end{pmatrix} 
\end{equation}
where $\sigma_k$ are the Pauli matrices, $\delta = \nu_{qb}-\nu_{RF}$ is the detuning between the pulse and the spin transition and $\Omega=g \mu_N B_{eff}/\hbar$ is the Rabi frequency. At resonance ($\delta = 0$), the state of the qubit will rotate around the x axis of the Bloch sphere at the frequency $\Omega$, resulting in a coherent oscillation between the population of the two states of the qubit. The visibility, defined as $V_{i j}=P_{i,j}+P_{j,i}$, as function of the pulse length for the three frequencies of the nuclear spin state ($\nu_1$=2.452GHz, $\nu_2$=3.128GHz, $\nu_3$=3.799GHz), are displayed in Figure 2(a). These oscillations exhibit a high fidelity coherent control of each nuclear spin transitions. As presented in Figure 2(d) and in details in Supplementary Materials (SM), the Rabi frequency fits linearly with the microwave amplitude, as theoretically predicted. The detuning  can also be adjusted by pitching the microwave pulse frequency. As shown in Figure 2(b-c), the higher the detuning, the lower the oscillation visibility and the higher the oscillation frequency. These results show that the molecular magnet single nuclear spin transistor geometry is a three qubits system, where all the dynamic parameters can be tuned. Furthermore, we measured coherence times of the order of a millisecond (see SM), which allow us to make more than thousand coherent spin manipulation before decoherence processes set in. \\

\section{III. Grover algorithm implementation}
In order to benefit from quantum parallelism and to implementation the Grover's research algorithm, the different transitions were driven simultaneously using a multi-chromatic microwave pulse. We make use of the generalized rotating frame \cite{Leuenberger2003} to treat the interaction of a multi-level system with a multi-chromatic pulse. Making the assumption of a near resonance condition for each pulse frequency and the rotating wave approximation, the Hamiltonian $H_{qd}$ of a four-state qudit system driven by a pulse composed of three frequencies is:

\begin{equation}
H_{qd}=\pi \hbar \begin{pmatrix}
0 & \Omega_1 & 0 & 0 \\
\Omega_1 & 2 \delta_1 & \Omega_2 & 0 \\
0 & \Omega_2 & 2 \delta_2 & \Omega_3 \\
0 & 0 & \Omega_3 & 2 \delta_3
\end{pmatrix}
\end{equation}

where $\delta_n$ are the frequency detunings of the n$^{th}$ transition and $\Omega_n$ are the Rabi frequencies of the n$^{th}$ transition. In this frame, the Hamiltonian is time independent. All unitary operations can be described via the evolution operator: 

\begin{equation}
U=e^{-iH_{qd}t/\hbar}
\end{equation}

The first gate of Grover's algorithm, the Hadamard gate, creates a quantum database, \textit{i.e.} prepares the system in a coherent superposition of all the nuclear spin states:
\begin{equation}
\ket{\Psi (\tau)} = U(\tau) \ket{\Psi_i}=\frac{1}{\sqrt{N}}\sum_{n=0}^{N-1}\ket{n}
\end{equation}
where $\tau$ is the pulse length that creates the superposition and $\ket{\Psi_i}$ the initial state. In the generalized rotating frame formalism, it is mandatory to find a combination of 2N-1 parameters (N-1 for $\Omega_n$, N-1 for $\delta_n$, and the evolution time) that satisfies this equation. Note that both phase and population of all states must be equal. We found the parameters using a variance minimisation of population and phase. The desired pulse was synthetized using a 24 GHz sampling rate AWG. \\
In this section, the visibility is defined as $V_{i,j}=P_{i,j}$. First, we applied the Hadamard gate to 2 states yielding a 2-state coherent superposition. Whereas a $\pi/2$ pulse creates a superposition of 2 states with a phase difference of $\pi$, the Hadamard gate with a detuning equal to the transition rate ($\delta= \Omega$) ensures a state population and phase equality. As shown in Figure 3(a), this is obtained starting from the 2$^{nd}$ state and driving the 2$^{nd}$ transition with the set of parameter $\delta_2$=$\Omega_2$= 3.1MHz and a pulse length of 115 ns. Next, a 3-state coherent superposition is displayed in Figure 3(b), starting from the 2$^{nd}$ state and driving the 1$^{st}$ and 2$^{nd}$ transitions, using $\delta_1$= $\Omega_1$=$\Omega_2$=2.4 MHz. Because of the Hamiltonian's symmetry, the 1$^{st}$ and the 3$^{rd}$ state have the same dynamics. Finally, for the 4-state superposition, we started with the 3$^{rd}$ state. The set of parameters $\delta_1$ = $\delta_2$ = $\delta_3$ = 0; $\Omega_1$ = 2.1MHz; $\Omega_2$ = 4.2MHz; $\Omega_3$ = 3.1MHz yielded a coherent superposition with a pulse length of 140 ns (Figure 3(c)). However, for this coherent superposition, each state has a different phase. \\

\begin{figure}
	\begin{center}
		\includegraphics[width=0.45\textwidth]{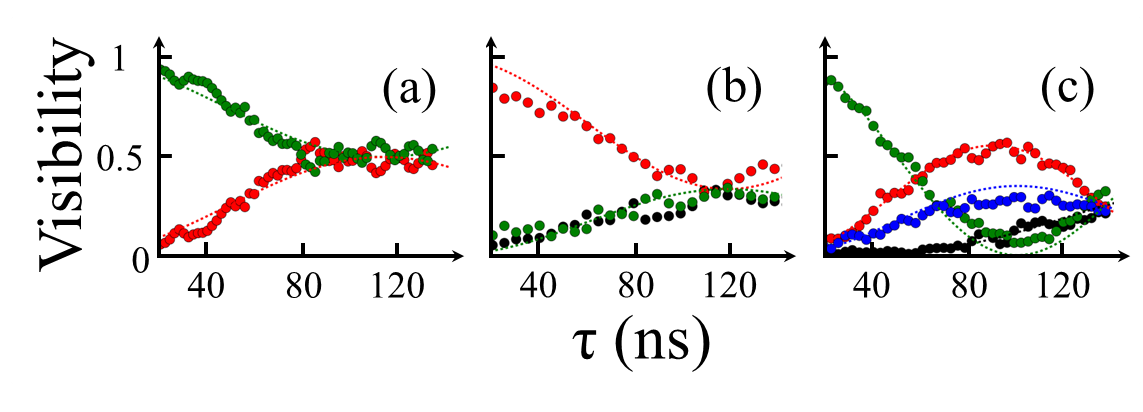}
		\caption{\label{fig3} Evolution of the nuclear spin states population as a function of the microwaves pulse length. The colour code is the same as used for Fig. 1. From top to down we show that we are able to create coherent superposition of 2, 3 and 4 single nuclear spin states. For the 2 (a) and 3 (b) states superposition we choose experimental parameters that ensure the same phase for all the states when they have the same population, leading to the so-called Hadamard gate. 
		}
	\end{center}
\end{figure}

These measurements show that this system can be used as a 4 states quantum directory. The second gate of Grover's algorithm is used to amplify the population of a researched state. This can be achieved by creating a resonant condition in between the superposed state and the researched state. Under this condition, the system will oscillate between these two states and after half-period of oscillation, will be in the researched state. Note that this unitary evolution has a $\sqrt{N}$ dependence of the period: 
\begin{equation}
\tau=\frac{\sqrt{N}}{4\Omega}
\end{equation}

as detailed in the SM. Experimentally, this resonant condition is obtained by applying a specific energy to the researched state. In the rotating frame, this means a specific detuning. In the general case of N-element database, the resonant condition is :
\begin{equation}
\braket{s| H_{qd} |s} = \frac{1}{N} \sum_{n,m} \braket{m | H_{qd} |n}
\end{equation}
where $\ket{s}$  is the researched state. For 3 states, this expression leads to:
\begin{equation}
\delta_s=\frac{\Omega_1+\Omega_2+\delta_1+\delta_2}{3}
\end{equation}
If we apply a detuning only to the researched state, the condition is:
\begin{equation}
\delta_s=\frac{\Omega_1+\Omega_2}{2}
\end{equation}
In order to synthesize the microwave pulse sequence with amplitude and phase control, we use a 24 GHz sampling rate AWG. Starting from the second state, the Hadamard gate is first applied to create the 3 states coherent superposition as presented in Figure 3(b). Then, the second pulse is generated with the same power (same $\Omega$) but with a modification of the frequency in order to satisfy the resonant condition. The pulse parameters (-$\delta_1$ = -$\delta_2$ = $\Omega_1$ = $\Omega_2$ = 3.4 MHz) select the 1$^{st}$ state, ($\delta_1$= $\Omega_1$ = $\Omega_2$ = 3.0 MHz) the 2$^{nd}$ state and ($\delta_2$ =  $\Omega_1$ = $\Omega_2$ = 4.9 MHz) the 3$^{rd}$ state (Figure 4 (a), (b) and (c) respectively). In all the cases, the population of the nuclear spin is clearly in the researched states (respectively a visibility of 0.9, 0.7 and 0.75). In order to underline the resonant character of this algorithm, we present a map of the visibility of the 3$^{rd}$ state as the function of the two frequencies that composed the pulse for a constant pulse length corresponding to the half-Grover period (Figure 4 (d)). As expected, the visibility is maximised when the resonant condition is reached ($\delta_1$ = 0 MHz and $\delta_2$ = $\Omega_1$ = $\Omega_2$ = 1.9 MHz). The experimental map compares well with the simulation (Figure 4(e)), demonstrating the implementation of the Grover search algorithm obtained using the resonance between a single state and a three superposed state of a single nuclear spin. \\

\begin{figure}
	\begin{center}
		\includegraphics[width=0.45\textwidth]{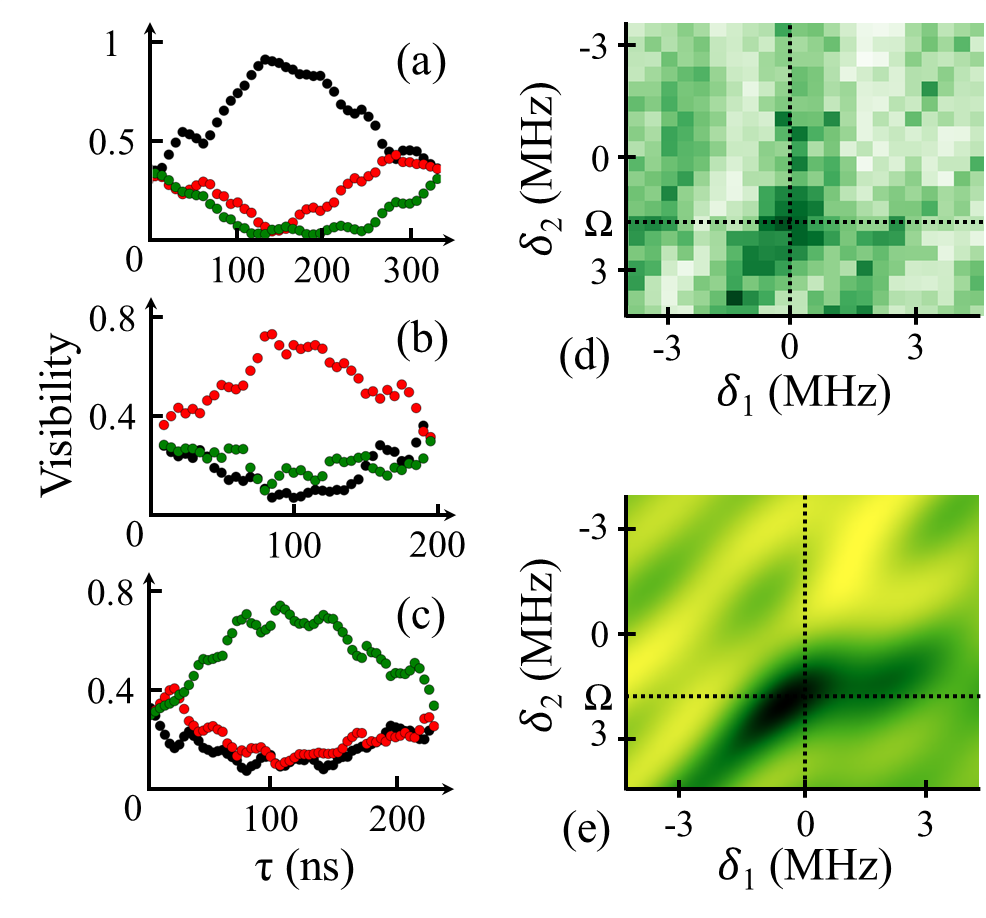}
		\caption{\label{fig4} Grover. Dynamic of the population in function of the unitary evolution pulse length. Starting from a superposed state (obtain by a Hadamard pulse sequence presented in Figure 3 (b)) we show that we are able to create an oscillation between this superposed state and a desired state. Depending on the detuning we choose to apply in this sequence, the population from either the black (a), the red (b) or the green (c) state increase. The Hadamard gate follows by this unitary evolution represent the implementation of the Grover's algorithm. Experimental (d) and theoretical (e) visibility as the function of the two pulse frequencies at fix pulse duration. The visibility is maximised when the resonant condition is satisfy. 
		}
	\end{center}
\end{figure}

These results show how the coherent control over a single nuclear spin embedded in a molecular spin transistor can be gained and read-out non-destructively. It leads to the first experimental implementation of the Grover's algorithm using a multi-levels system. The presented two-step quantum operation can be extended to be performed on alternative spin Qubit devices. The great diversity of available molecular magnets with their inherent tunability will potentially provide higher nuclear spin values that might make accessible much bigger databases for the field of molecular quantum computation. \\

We gratefully acknowledge E. Eyraud, D. Lepoittevin, and C. Hoarau for their technical contributions and motivating discussions. We thank T. Fournier, T. Crozes, B. Fernandez, S. Dufresnes and G. Julie for Nano-fabrication development, E. Bonet and C. Thirion for help with software development, R. Vincent, S. Thiele for the development of the experiment. Samples were fabricated in the NANOFAB facility of the N\'eel Institute. This work is partially supported by ANR-13-BS10-0001 MolQuSpin. \\

\end{document}